\documentstyle[pre,aps,epsf, eqsecnum, multicol]{revtex}

\def\top#1{\vskip #1\begin{picture}(290,80)(80,500)\thinlines \put(
65,500){\line( 1, 0){255}}\put(320,500){\line( 0, 1){
5}}\end{picture}}
\def\bottom#1{\vskip #1\begin{picture}(290,80)(80,500)\thinlines \put(
330,500){\line( 1, 0){255}}\put(330,500){\line( 0, -1){
5}}\end{picture}}

\begin{document}
\draft
\title{Boojums and the Shapes of Domains in Monolayer Films}
\author{Jiyu Fang, Ellis Teer and Charles M. Knobler}
\address{Department of Chemistry, UCLA\\ Los Angeles, California 90095-1569}
\author{  Kok-Kiong Loh and Joseph Rudnick }
\address{Department of Physics, UCLA \\ Los Angeles, California 90095-1547}
\date{\today}
\maketitle
\begin{abstract}
Domains in certain  Langmuir monolayers support a texture that is the
two-dimensional version of  the  feature known as a boojum. Such a texture has a
quantifiable effect on the shape of the domain with which it is associated. The
most noticeable consequence is a cusp-like feature on the domain boundary. We
report the results of an experimental and theoretical investigation of the shape
of a domain in a Langmuir monolayer. A further aspect of the investigation is
the study of the shape of a ``bubble'' of gas-like phase in such a monolayer.
This structure supports a texture having the form of an {\em inverse} boojum.
The distortion of a bubble resulting from this texture is also studied. The
correspondence between theory and experiment, while not perfect, indicates that
a qualitative understanding of the relationship between textures and domain
shapes has been achieved.
 \end{abstract}
\begin{multicols}{2}
\section{Introduction}
The direct observation of monolayers at the air/water interface by the
methods of polarized fluorescence microscopy (PFM)\cite{PFM} and Brewster
angle microscopy (BAM)\cite{BAM} reveals that the films possess complex
textures similar to those observed in liquid crystals. The textures are
generally found in ``tilted'' phases, i.e. in phases in which the long axes
of the molecules in the film are are not perpendicular to the water surface
but  are uniformly tilted with respect to the normal. The textures are the result
of the spontaneous organization of the molecular tilt azimuth on macroscopic
length scales. They can be understood, at least qualitatively, in terms of
continuum elastic theories of smectic liquid crystals\cite{smectics}. 

Many striking textures have been found in domains of condensed tilted phases,
such as the $L_2$ phase, that are surrounded by an isotropic phase, either liquid
[often designated $LE$, (liquid-expanded) or $L_1$] or gas ($G$)\cite{review}.
Among those textures are boojums, in which the tilt azimuth varies continuously
and appears to radiate in some cases from a defect located at the edge of the
domain and in others from a ``virtual'' defect in the isotropic phase\cite{L&S}.
The shapes of domains that contain the boojum texture are not round; they have a
cusp-like deformation pointing toward the defect. 

In essence, the equilibrium shape of a domain containing a boojum is like the
equilibrium shape of a crystal, which can be calculated with the use of the
Wulff construction\cite{Wulff}. In that procedure it is assumed that the
surface energy of the crystal is determined solely by the anisotropy of the
bulk energy. For a monolayer domain containing a \underline{nontrivial} texture 
this is not the case; the boundary---more properly, line---energy is comparable
to the energy associated with the alignment of the tilt azimuth, so shape and
texture must be calculated self-consistently.

Such a calculation has been performed by Rudnick and Bruinsma\cite{RudBru},
who showed that there is a cusp on the domain boundary, quantified by an
``excluded'' angle $\Psi$ that varies with $R$, the radius of the domain.
\vadjust{\vskip 0.1in
\narrowtext
\begin{figure}
\makebox{\epsfbox{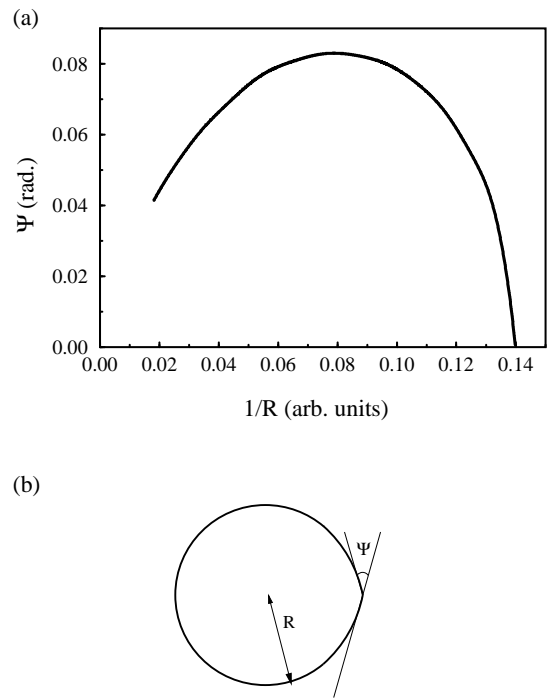}}
\caption{(a) The excluded angle, $\Psi$, plotted as a function of the 1/R. 
(b) The definition of the excluded angle $\Psi$.}
\label{R+B} 
\end{figure}
\vskip 0.05 in}
Their results are displayed concisely in Fig. \ref{R+B}, which
contains a pictorial definition of $\Psi$ and a plot of $\Psi$ versus $1/R$. 
The excluded angle is equal to zero when the
domain is circular and increases as the cusp sharpens. 
Rudnick and Bruinsma predict that $\Psi$ vanishes in the limit $R=\infty$ and
that the angle is initially linear in $R^{-1}$. The excluded angle goes through a
maximum and then decreases to zero for sufficiently small domains. Schwartz,
et al.\cite{Schwartz} made PFM measurements of $\Psi$ for domains of
pentadecanoic acid ranging in size from 12 to 120 $\mu$m. They  found
that the excluded angle increases with $1/R$, but roughly as a power law
with an exponent of 0.3 rather than unity. There was no evidence of a maximum.

Rivi\`{e}re and Meunier\cite{RivMeun} carried out BAM studies of condensed
domains of hexadecanoic acid surrounded by a gaseous phase and measured the
distance, $d$, of the virtual defect from the domain edge as a function of $R$.
By assuming that the deformation of the domain is small, they were able to
derive a relation between the quantity $\rho=d+R$ and the domain radius from
continuum elastic theory. They compared the calculated and measured plots of
$\rho$ versus $R$ and were able to obtain values for the ratio of the
bend/splay elastic constant to the anisotropic part of the line tension.
They also made a qualitative comparison between the measured elongation of
the boojum-containing domains and the elongation calculated from a
generalized Gibbs-Thompson equation that was obtained by Galatola and
Fournier\cite{GalFour}, this under the assumption that the texture is that
of an undistorted boojum. 

Experimental tests of the theory have been limited by the small range of domain
sizes that could be examined by optical microscopy. In this paper we report new
measurements of the excluded angle for much smaller domains. This has been
accomplished by transferring the monolayers to a solid support by the
Langmuir-Blodgett technique and obtaining images with the use of scanning force
microscopy. We also describe measurements on ``inverse'' boojums, which occur
when there is a ``bubble'' of isotropic phase imbedded in a tilted phase. A new
theoretical analysis of the boojum problem, which is an extension and
reformulation of the treatment of Rudnick and Bruinsma, is then described.
Comparisons are made between the experiments and this theory.

\section{experiment}

Monolayers of pentadecanoic acid (Nu-Chek Prep, $\ge$ 99\%) were deposited
from chloroform (Fisher spectranalyzed) solutions onto water (Millipore
Milli-Q, pH 5.5) in a NIMA Type 611 trough. They were transferred onto mica
substrates that had been freshly cleaved with adhesive tape and immediately
inserted into the water subphase. The transfers were performed by
withdrawing the mica from the subphase at constant speeds ranging from 0.5 to 2
mm/min. Boojums of the $L_2$ phase surrounded by the $LE$ phase were obtained
in the coexistence region at temperatures between 15 and 25$^{\circ}$C;
boojums of the $G$ phase surrounded by the $L_2$ phase were obtained in
transfers at 7$^{\circ}$C.

The transferred monolayers were imaged with a scanning force microscope (SFM)
(Park Scientific Instruments) using a 100-$\mu$m scanner at room temperature
in ambient atmosphere. The instrument was calibrated on the micrometer scale
by imaging grids with known spacings and on the nanometer scale with mica
standards. A microfabrication triangular Si$_3$N$_4$ cantilever with a
normal spring constant of 0.05 N/m was used for the measurements. The vertical 
bending and lateral torsion of the cantilever were monitored by reflecting a
laser bean from the end of the cantilever into a four-segment photodetector so
that the topographic and frictional force images of the samples could be
obtained simultaneously and independently of each other. All images were
obtained in the constant-force mode. The loading forces ranged from 1 to 5 
$\mu$N.

Brewster angle microscope images of Langmuir monolayers were obtained with an
instrument of our own construction. Details of the instrument and the
experimental procedures can be found elsewhere\cite{Fischer}.

\section{experimental results}

The key to the success of the experiment is the requirement that the boojums
survive the transfer process and that their shape on the support be the same
as that on the water surface. It has often been observed that  the
continuous $LE$ phase breaks up into small islands when it is transferred to a
solid support. The behavior has been found in monolayers of stearic acid
deposited on thin polyethyleneimine films\cite{polyeth},
dimyristolyphosphatidic acid deposited on glass\cite{dimyr} and
octadecyltrichlorosilane on acid-treated mica\cite{octa}. However, as can be
seen
in Fig. \ref{domainpicture}(a), despite the  breakup of the $LE$ phase, the boojum shapes of the $L_2$ phase survive. LB
transfer is also known to cause flow in monolayers\cite{flow1,flow2}, and there
is evidence of this in the transfers made at 2 mm/min. As is evident in Fig.
\ref{domainpicture}(b), the domains are elongated in the dipping direction.
However, at 0.5 mm/min no distortion is evident and the boojums are not aligned
in the dipping direcion.

A plot of the excluded angle against the reciprocal of the domain diameter is
shown in Fig. \ref{domaindata1}. The same data are displayed in a log-log plot
in Fig. \ref{domaindata2}. In the range 10 - 20 $\mu$m the data from the SFM
studies agree with those obtained earlier by PFM\cite{Schwartz}. We note that
the slope of the log-log plot at large $R$ is less than unity, and that there 
is a non-zero intercept in the $\Psi$ vs $1/R$ plot. The fact that the
intercept is not zero had been overlooked in the earlier work, in which the data
were presented on a log-log plot. By extending the measurements to smaller
\vadjust{\vskip 0.1in
\begin{figure}
\makebox[3in]{\epsfbox{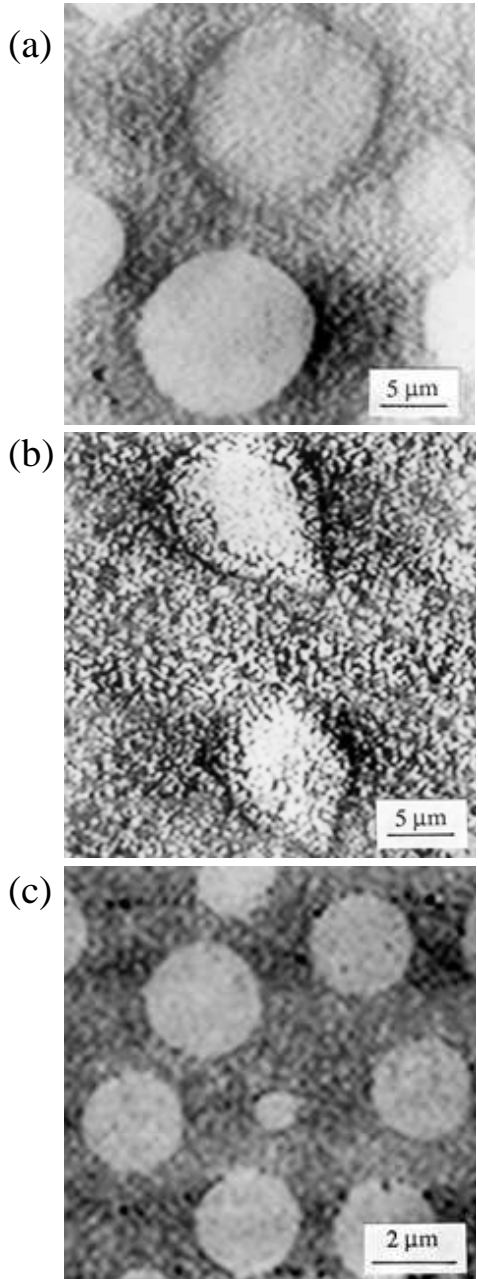}}
\caption{Atomic force microscope images of domains associated with
boojums. Monolayers were transferred onto mica from the $L_2-LE$ coexistence
region at a speed of 0.5mm/min (a),  (c), and at 2 mm/min (b).} 
\label{domainpicture}
\end{figure}
\vskip 0.05 in}
domains we have observed that there is a maximum and that the small domains
become circular. There is no evidence of a cusp in domains smaller than 2 $\mu$m
in diameter; see Fig. \ref{domainpicture}(c).

The good agreement between the PFM and SFM studies for the larger domains gives
us confidence that the shapes have not been seriously changed during the
transfer process or greatly affected by the change in substrate. We cannot
be certain that this is true of the smaller domains, but it does seem likely
that the effect of hydrodynamic flows and variations in the substrate will
\vadjust{\vskip 0.1in
\begin{figure}
\makebox{\epsfbox{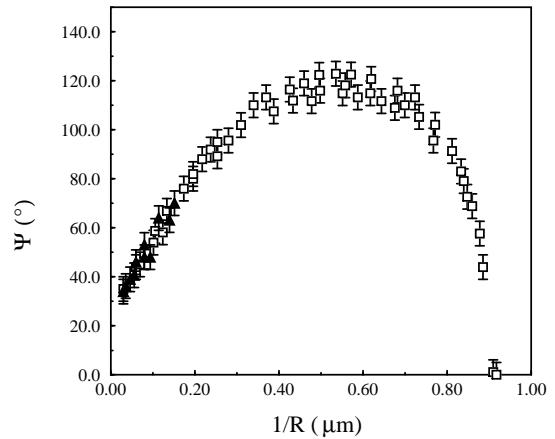}}
\caption{The excluded angle of the cusp-like feature on a domain supporting a
boojum texture, plotted against the reciprocal of the domain radius.  Filled triangles represent data for domains on water surface and empty squares represent data for domains tranferred onto solid support. }
\label{domaindata1}
\end{figure}
\begin{figure}
\makebox{\epsfbox{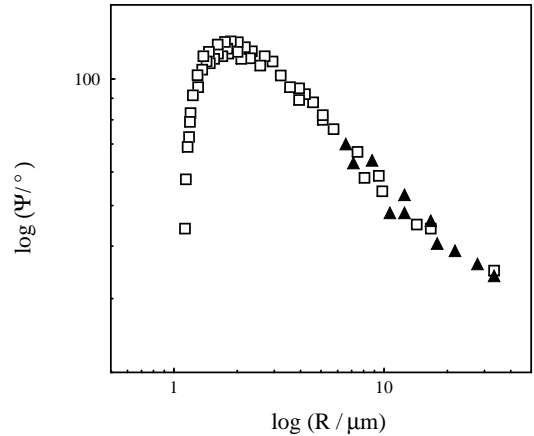}}
\caption{A log-log plot of the dependence on radius of
the excluded angle of the cusp-like feature on a domain supporting a boojum
texture.  Filled triangles represent data for domains on water surface and empty squares represent data for domains tranferred onto solid support.}
\label{domaindata2}
\end{figure}
\vskip 0.05 in}
be most serious in the larger domains.

\begin{figure}
\makebox[3in]{\epsfbox{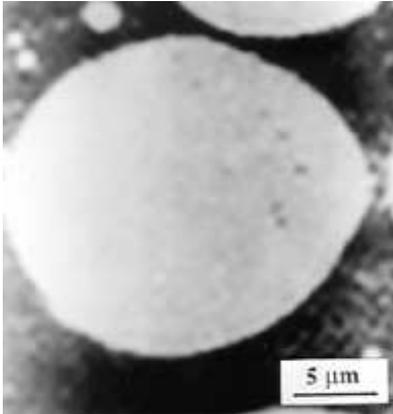}}
\caption{Frictional force image of an inverse boojum consisting
of a bubble of the isotropic $G$ phase surrounded by the anisotropic
$L_2$ phase. The monolayer was transferred onto mica from the $L_2-G$ 
coexistence region at a speed of 0.5 mm/min.}
\label{friction}
\end{figure}

In the PFM studies \cite{Schwartz} it was observed that inclusions
of the $LE$ phase in the $L_2$ domains also had cusps. When we examined films at
temperatures below the $LE$-$L_2$-$G$ triple point at 13$^{\circ}$C\cite{trip},
we found similar cusp-shaped regions in the isotropic $G$ phase surrounded by the
anisotropic $L_2$ phase. These domains also survive transfer to a solid
support.  Fig. \ref{friction} is a frictional force image of such an ``inverse''
boojum. The frictional force within the ``inverse'' boojum is high because the
tip images the mica substrate and not the dilute $G$ phase.  This is  also
confirmed by the topographic image, which was obtained simultaneously.  The
height profile shows that the inverse boojum extends to the substrate surface. 

\begin{figure}
\makebox{\epsfbox{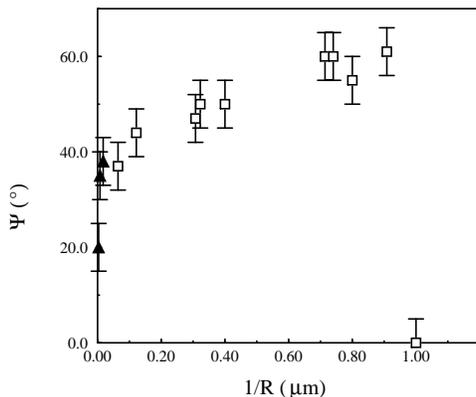}}
\caption{The excluded angle of the cusps in bubbles of isotropic $G$ phase
surrounded by anisotropic $L_2$ phase, plotted versus the reciprocal of the
bubble radius. filled triangles represent data for bubbles on water surface and empty squares represent data for bubbles tranferred onto solid support. Note the similarity with Fig. 3.}
\label{bubbledata1}
\end{figure}

Measurements of the excluded angles in these inverse boojums as a function of
their radii are plotted in Fig. \ref{bubbledata1}. The behavior is similar to
that found for the ordinary boojums; the excluded angle has a nonzero intercept,
increases roughly as $R^{-0.3}$, goes through a maximum and falls to zero.
Included in the figure are three points obtained from BAM studies of inverse
boojums in Langmuir monolayers of pentadecanoic acid. There is a reasonable
agreement with the SFM experiments for large domains. However, cusp angles are
difficult to measure in small domains by BAM. They can be more easily and
precisely measured by fluorescence microscopy, but the precipitation of the
probe in the $L_2$-$G$ two-phase region made this impossible.

\section{The inverse boojum}

The boojum texture is an outcome  of the competition between bulk and
boundary energies in a bounded domain containing a condensed phase.
When the boundary energy has the particular anisotropic form displayed below,
and the domain is a perfect circle, this texture is an exact solution
to the energy minimization equations\cite{RudBru}. At the same time as the
$XY$-like order parameter adjusts in response to the imperatives of energy
minimization, the boundary also distorts so as to accomodate the now
inhomogeneous environment presented by the texture. It has been found that when
the energy is as given by Eqs. (\ref{hamil}) and (\ref{simplesig}), a circular
domain satisfies the extremum equations. 

This means that the simple version of the energy of a domain in a Langmuir
monolayer leads to a liquid-condensed-phase domain supporting a nontrivial
texture whose boundary is perfectly circular. To explain the domain shapes
observed in the experiments reported here and previously \cite{Schwartz,RivMeun}
it is necessary to introduce elaborations on the system's energy. The addition of a higher-order harmonic term to the line
tension does give rise to a
distortion of the boundary \cite{RudBru} and a feature in the form of a cusp appears. The theory is in qualitative accord
with the experiments except in the limiting large-domain behavior,  where it predicts that the cusp angle vanishes rather
than going to a limiting non-zero value as observed for both domains and bubbles.

The shape of bubbles of an isotropic phase inside an ordered phase was not considered in the theory. When there is a
two-dimensional bubble of gas phase inside an extended region
of the liquid-condensed phase a texture appears in the liquid-condensed phase
that is closely related to the virtual boojum observed in two-dimensional
liquid-condensed droplets. This texture minimizes the total energy
$H\left[\Theta(x,y) \right]$, given by
\begin{eqnarray}
H\left[\Theta(x,y) \right] =&& \int_{{\rm bulk}}\frac{\kappa}{2} \left|
\vec{\nabla}\Theta \right|^2\,dxdy \nonumber \\ 
&& + \int_{{\rm boundary}}\sigma\left(\theta -
\Theta(x,y)\right)\,ds,  \label{hamil}
\end{eqnarray}
where $\Theta(x,y)$ is the angle between the vector order parameter $\hat{c}$
and the $x$-axis, while $\theta$ is the corresponding angle for the unit
normal, $\hat{n}$. The coefficient $\kappa$ is the Frank, or ``spin stiffness''
constant. The energy above follows from the assumption of equal bend and splay
moduli, which is known to represent, at best, an approximation to physical
reality. The function $\sigma\left(\theta - \Theta(x,y)\right)$ is the surface
energy, which depends  on the relative orientations of $\hat{c}$ and the unit
normal to the boundary, $\hat{n}$. The extremum equations that arise are  
\begin{eqnarray}
\nabla^2\Theta(x,y) &=& 0   \label{bulk_eqn} \\
\kappa \frac{\partial \Theta}{\partial n} + \sigma'(\theta - \Theta) &=& 0
\label{bound_eqn}
\end{eqnarray}
\vadjust{\vskip 0.1in
\begin{figure}
\makebox[3in]{\epsfbox{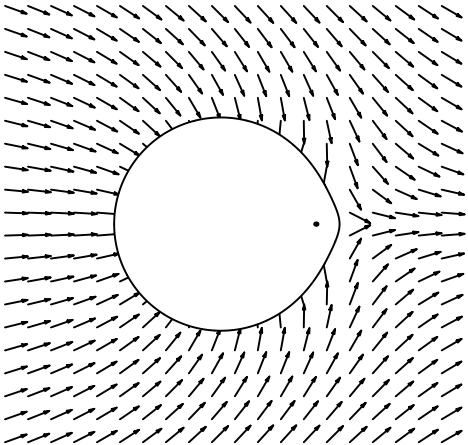}}
\caption{The texture associated with an inverse  boojum and the position of the
virtual singularity, given by Eqs. (\ref{invsbooj}) and (\ref{boojloc}). The
distortion of the bubble from a  circle, obtained from Eq.
(\ref{polar5}),  is also indicated on the Figure.  The parameters used are
$\kappa/a=0.4\mu m$, $a/\sigma_0=0.16$ and $R_0=2.4 \mu m$.}
\label{virtualboojum}
\end{figure}
\vskip 0.05 in}
Eq.(\ref{bulk_eqn}) applies outside the bubble and Eq.(\ref{bound_eqn}) at the interface. If
$\sigma(\phi)$ has the form
\begin{equation}
\sigma(\phi) = \sigma_0 + a \cos \phi ,   \label{simplesig}
\end{equation}
and the bubble is a perfect circle,
then a solution for $\Theta(x,y)$ that satisfies the extremum equations
Eq. (\ref{bulk_eqn}) and Eq. (\ref{bound_eqn}) is
\begin{equation}
\Theta(x,y) = \frac{1}{i}\left[ \ln\left(1 - \frac{\alpha}{x-iy} \right) - \ln
\left( 1 - \frac{\alpha}{x+iy} \right) \right] .  \label{invsbooj}
\end{equation}
Given the similarity between the mathematical structure of the right-hand
side of Eq. (\ref{invsbooj}) and the expression for  $\Theta(x,y)$ in the case of
the virtual boojum (see ref. \cite{RudBru}) this solution can be characterized as
an \underline{inverse} boojum. It is a texture with a singularity that lies a
distance $R_B$ from the origin, where 
\begin{eqnarray} \frac{R_B}{R_0} &=&
\frac{\alpha}{R_0} \nonumber \\ 
&=& \frac{aR_0/\kappa}{\sqrt{1 + \left(
aR_0/\kappa \right)^2}+1}. \label{boojloc} 
\end{eqnarray}
The above relation serves to fix the value of the quantity $\alpha$. Note that
$R_B$ is always less than the radius of the bubble. The singularity lies
within the bubble. The texture associated with a virtual boojum is displayed in
Fig. \ref{virtualboojum}.

\begin{figure}
\makebox[3in]{\epsfbox{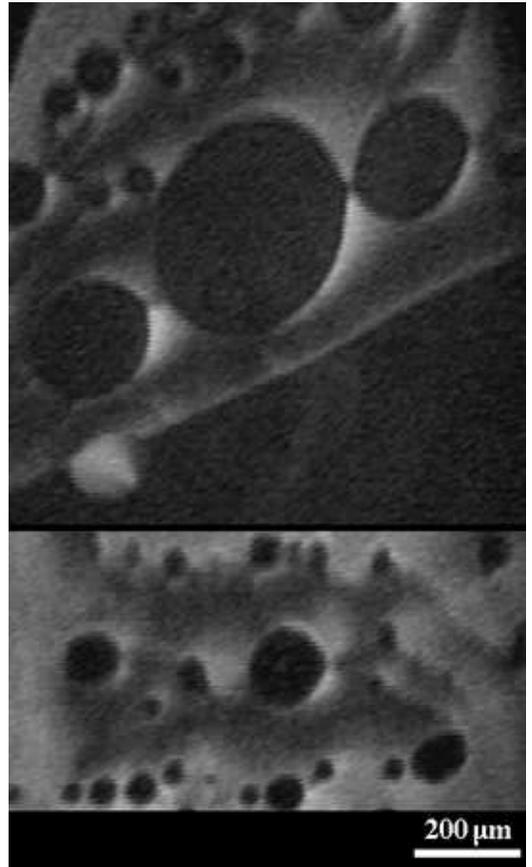}}
\caption{BAM images of inverse boojums in Langmuir monolayers of pentadecanoic
acid at $5^{\circ}$C.}
\label{actualBAM}
\end{figure}

When a virtual boojum exists in a condensed-phase domain,  the orientation of the 
director leads to an optical anisotropy that should be observable by
Brewster-angle microscopy\cite{Schwartz,RivMeun}. A ``sunburst''
\vadjust{\vskip 0.1in
\begin{figure}
\makebox{\epsfbox{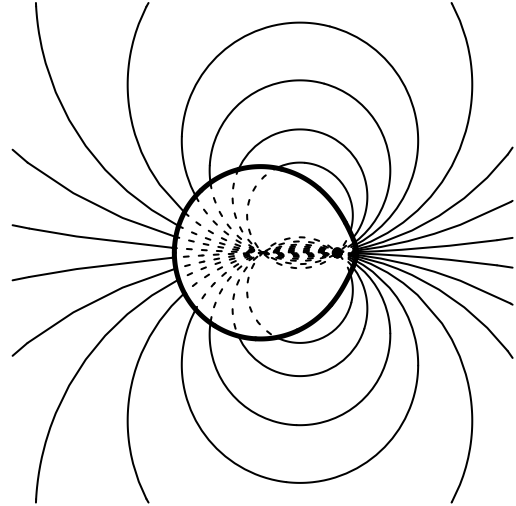}}
\caption{The curves along which the order parameter points in a fixed direction
when there is an inverse boojum texture in the immediate vicinity of a bubble.The 
parameters used are  $\kappa/a=0.4\mu m
$, $a/\sigma_0=0.16$ and $R_0=2.4 \mu m$.}
\label{Brewster} 
\end{figure}
\vskip 0.05 in}
of darker and lighter lines radiates from the point of singularity. Regions of 
constant intensity on the
pattern correspond to portions of the texture in which the order parameter
points in a fixed direction. In the case of the inverse virtual
boojum, there should be an analogous signature of the texture in the surrounding condensed phase.  Using Eq.
(\ref{invsbooj}) one finds
straightforwardly that the orientation of the order parameter is constant
along the perimeter of a circle. This means that points of equal orientation
fall on circular countours as shown in Fig. \ref{Brewster}. The contours appear
to pass through two points inside the bubble. One of the points is the
location of the singularity, and the other is the center of the bubble. The
optical anisotropy that will be observed by BAM depends on the molecular
parameters and the orientation of the boojum with respect to the direction of
the incident light. Qualitative simulations of the images are shown in Fig.
\ref{simulatedBAM}. They have been obtained  by using the relation between tilt
azimuth and  reflectivity calculated by Tsao, et. al.\cite{Tsao}. These images
can be compared with the BAM images in Fig. \ref{actualBAM}, which show similar
variations in the reflectivity. It is interesting to note that the boojums
appear to align. A similar alignment of isotropic droplets in a nematically
ordered host phase has been observed\cite{Poulin}.
This alignment is attributed to a long-range interaction produced by
distortions in the director fields.

\section{Bubble Shape: Theory }
In equilibrium a bubble takes on the shape that minimizes  the system's free
energy. If the interior is uniform and the line tension of an element of
\vadjust{\vskip 0.1in
\begin{figure}
\makebox{\epsfbox{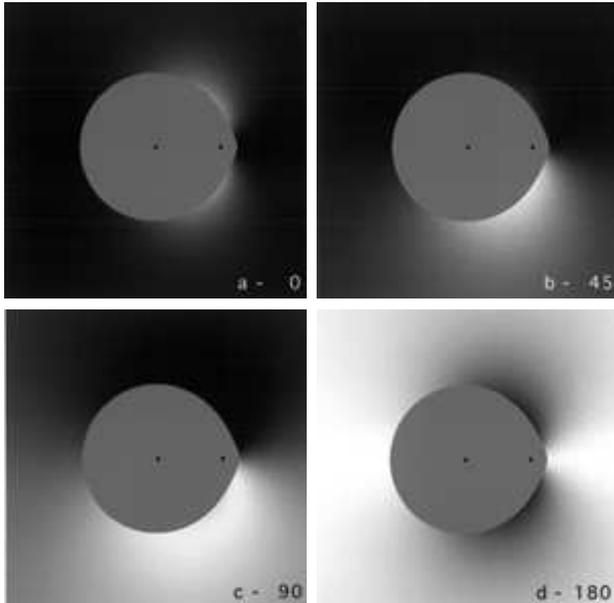}}
\caption{Simulated BAM images of inverse boojums for different orientations
with respect to the incident light. The orientation of the beam with respect to
the symmetry axis of the boojum is specified in the lower corner of each
figure.} 
\label{simulatedBAM} 
\end{figure}
\vskip 0.05 in}
boundary depends  on the  orientation of the boundary element, the Wulff
construction can be utilized to yield the shape. In the case of inhomogenous
interior structure, a generalized version of the Wulff construction provides an
approach to the solution\cite{RudBru}. This method applies to the bubble in much
the same way as it does to a
domain.  One discovers immediately that a circular bubble does {\em not}
minimize the system energy of the form (\ref{hamil}) with $\sigma(\phi)$ given by
Eq. (\ref{simplesig}), in contrast to the result obtained for a domain
\cite{RudBru}, for which a circle represents the minimum energy domain shape.
However, the method cannot be applied directly because the curvature of the
boundary changes sign. As a consequence, the radius of curvature of the
boundary may become infinite at points along it, and this proves awkward in the
context of the Wulff construction. However, a reparameterization allows one to
proceed with the analysis. We find that there is no feature that can, strictly
speaking, be described as a cusp, in that the slope of the bounding curve
has no discontinuity. There is, on the other hand, a feature on the boundary
whose shape can be quantified in terms of an excluded angle.

We have utilized an  alternative approach to the determination of the
bubble's shape, to provide a check on our Wulff construction analysis.
We  adopt polar coordinates.  This formulation confirms the results of the
analysis based on the Wulff construction. We obtain explicit results for the
bubble's shape. Cusp angles are determined both analytically and by means of
measurements on figures generated with the use of explicit analytical results,
and these angles provide the basis for comparison with experiment.

The remainder of this section consists of a review of the generalized Wulff
construction, the development of the  alternative approach and a
summary of results. Details are deferred to a forthcoming
publication\cite{forthcoming}.

\subsection{Generalized Wulff construction}

Following the approach of Rudnick and Bruinsma, we attempt to apply
the generalized Wulff construction \cite{RudBru} to determine the shape of the
bubble.  In our application,  this technique is used to
minimize the total energy Eq. (\ref{hamil}) with respect to the Burton, Cabrera and
Frank (BCF) parameterization of bubble shape,  i.e. $R(\theta)$ and $\theta$
\cite{Burton}.

We rederive the Wulff construction so as to remove the approximation---utilized
in earlier work\cite{RudBru}---that the radius of curvature, $R+R''$,  is
constant. We obtain 
\begin{equation}
\frac{\delta}{\delta R}  \int_{{\rm bulk}}\frac{\kappa}{2} \left|
\vec{\nabla}\Theta \right|^2\,dxdy
=-\frac{\kappa}{2}\left|\vec{\nabla}\Theta\right|^{2}\left[R+R^{\prime\prime}\right],
\label{Wulff1}
\end{equation}

It can be shown that if $\sigma(\phi)=\sigma(-\phi)$,  we have for the bubble
\begin{eqnarray}
&&\frac{\delta}{\delta R} \int_{{\rm boundary}}\sigma \left(\theta -
\Theta(x,y)\right)\,ds \nonumber \\
&&=\sigma +\sigma^{\prime\prime}-(R+R^{\prime\prime})\left[\sigma'(\cos \theta
\frac{\partial}{\partial
x}+ \sin  \theta\frac{\partial}{\partial y}) \right. \nonumber \\
&&\left. +\sigma''(-\sin \theta
\frac{\partial}{\partial x} + \cos \theta \frac{\partial}{\partial y})
\right] \Theta, \label{Wulff2}
\end{eqnarray}

Adopting (\ref{simplesig}) and (\ref{invsbooj}),  the equation for the shape
of the bubble is given by

\end{multicols}
\widetext
\top{-2.8cm}
\begin{eqnarray}
\lefteqn{\frac{\delta H}{\delta R}=0\Rightarrow} \nonumber \\
&&r+r^{\prime\prime}= \nonumber \\
&&\frac{(1-2\alpha \cos\theta+\alpha^{2})^2}{2\alpha(\alpha+\delta)\cos 2
\theta -4\alpha(1+\alpha^{2}+2\delta\alpha)\cos \theta+1+4\alpha^{2}+
\alpha^{4}+\delta\alpha(1+6\alpha^{2}-\alpha^{4})},   \label{Wulffx}
\end{eqnarray}
\bottom{-2.7cm}
\begin{multicols}{2}

\hspace{-.15in}where $\delta=a/\sigma_{0}$, $R=R_0 r$ and $R_0$ is the radius of the
unperturbed circle. 
The denominator of Eq. (\ref{Wulffx})  passes through zero in regimes of interest, 
and this complicates the analysis. It proves useful to reparameterize the
system in Cartesian coordinates.

The relations between the parameterization using Cartesian coordinates
and the BCF parameterization are as follows:
\begin{eqnarray}
x&=&r\sin \theta + r^{\prime}\cos \theta, \label{WulffCartp1}\\ 
y&=&r\cos\theta - r^{\prime}\sin\theta. \label {WulffCartp2}
\end{eqnarray}
Using Eq. (\ref{WulffCartp1}) and Eq. (\ref{WulffCartp2}), we  have the following
relations for nearly circular boundaries and in the region where $\theta$  is
small 
\begin{eqnarray} \frac{1}{r+r^{\prime\prime}}&=& 
-\frac{y^{\prime\prime}}  {\sqrt{1+y^{\prime 2}}^{3}} \approx -y^{\prime\prime}
\label{WulffCartp5}\\ 
\cos \theta &\approx& \sqrt{1-x^2} \approx
1-\frac{x^{2}}{2}-\frac{x^{4}}{8} \label {WulffCartp6} 
\end{eqnarray} 
Substituting Eq. (\ref{WulffCartp5}) and
Eq. (\ref{WulffCartp6}) into  Eq. (\ref{Wulffx}), we obtain
\begin{equation}
y^{\prime\prime}=-\frac{C_{2}(x^{2}-k_{+})(x^{2}-k_{-})}{[\alpha x^{2}+
(1-\alpha)^{2}]^{2}},   \label{Wulff3}
\end{equation}
where
\begin{eqnarray}
k_{\pm}&=&\frac{-C_{1}\pm\sqrt{C_{1}^{2}-4C_{0}C_{2}}}{2C_{2}},
\label{Wulff4} \\
C_{2}&=&\frac{\alpha}{2}(1+\alpha^2+2\delta\alpha),
\label{Wullf5} \\
C_{1}&=&2\alpha(1-\alpha)^{2}-4\delta\alpha(1-\alpha),   \label{Wulff6} \\
C_{0}&=&(1-\alpha)^{4}+\delta\alpha(3+\alpha)(1-\alpha)^{3}.    \label{Wulff7}
\end{eqnarray}
The integration of Eq. (\ref{Wulff3}) is readily performed easily to yield
\begin{eqnarray}
&& y^{\prime}= \nonumber \\
&-&\frac{C_{2}x}{\alpha^{2}}\nonumber \\
&-&\frac{C_{2}(1-\alpha)^{4}-C_{1}\alpha
(1-\alpha)^2+C_{0}\alpha^{2}}{2\alpha^{2}(1-\alpha)^{2}}\frac{x}
{(1-\alpha)^{2}+\alpha x^{2}} \nonumber \\
 &+&\frac{3C_{2}(1-\alpha)^{4}-C_{1}(1-\alpha)^2-C_{0}\alpha^{2}}
{2(1-\alpha)^{3}
\alpha^{\frac{5}{2}}}\tan^{-1}\frac{\alpha^{\frac{1}{2}}x}{1-\alpha}\nonumber \\
&+& K_{y^{\prime}}, \label{Wulff8}
\end{eqnarray} 
and
\begin{eqnarray}
y&=&-\frac{C_{2}x^{2}}{2\alpha^{2}}-\frac{2(1-\alpha)^2+C_{1}\alpha}
{2\alpha^{3}}
\ln[\alpha x^{2}+(1-\alpha)^2] \nonumber \\
&&+\frac{3C_{2}(1-\alpha)^{4}-C_{1}(1-\alpha)^{2}-C_{0}\alpha^{2}}
{2(1-\alpha)^{3}\alpha^{\frac{5}{2}}}x\tan^{-1}
\frac{\alpha^{\frac{1}{2}}x}{1-\alpha}  \nonumber \\
&&+K_{y^{\prime}}x+K_{y}.    \label{Wulff9}
\end{eqnarray}

The next step is to fix the constants of integration $K_{y^{\prime}}$ and $K_y$.
Symmetry requires $y^{\prime}(-x)=-y^{\prime}(x)$, and yields
$K_{y^{\prime}}=0$.    The bounding curve $y(x)$ will merge into a circle of
radius 1, centered at the origin.  The merging point, $(x_0,  y(x_0))$, is taken
to be the point at which $y^{\prime}(x_0)= -x_0/\sqrt{1-x_0^{2}}$. This leads to 
\begin{eqnarray} 
&&K_{y}=\nonumber \\
&&\frac{C_{2}x_0^{2}}{2\alpha^{2}}+\frac{2(1-\alpha)^{2}
+C_{1}\alpha}{2\alpha^{3}}\ln[\alpha x_0^{2}+(1-\alpha)^2] \nonumber \\
&&-\frac{3C_{2}(1-\alpha)^{4}-C_{1}(1-\alpha)^{2}-C_{0}\alpha^{2}}{2
(1-\alpha)^{3}\alpha^{\frac{5}{2}}}x_0\tan^{-1}\frac{\alpha
^{\frac{1}{2}}x_0}{1-\alpha}\nonumber \\
&&+\sqrt{1-x_0^{2}}, \label{Wulff10}
\end{eqnarray}

\begin{figure}
\makebox[3in]{\epsfbox{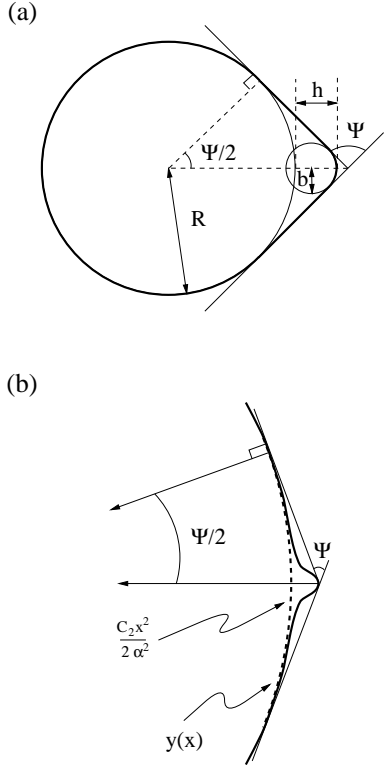}}
\caption{(a) The way in which $\Psi$ is defined where the boundary is smooth. This method leads
to an analytical expression for the excluded angle of a cusp-like feature. (b) For $R_0 \gg
\kappa/a$, the "cusp"-like feature show up differently and requires a new definition of $\Psi$.}
\label{excluded}
\end{figure}

The boundary of the bubble Eq. (\ref{Wulff9})  is smooth in that there is 
no discontinuity in the slope of the bounding curve.  For very small $R_0$, the 
radius of curvature of the boundary close to the virtual boojum,
$b=1/y^{\prime\prime}(0)$, is approximately equal to $R_0$. The bubble appears to
be circular and the excluded angle, $\Psi$, is zero. The ratio $\beta=b/R_0$
decreases as $R_0$ increases. This causes the bubble to sharpen at one end.  When
$\beta$ exceeds a critical value $\beta_c$, this sharper end  has properties in
common with be a cusp, and  one is able to derive a excluded angle, $\Psi$, as
depicted in Fig. \ref{excluded}a.  It can be shown that $\Psi$ is given by the
following: \begin{equation}
\Psi=2\cos^{-1}\frac{1}{1+\frac{h}{1-b}},  \label{Wulff11}   
\end{equation}  
where $h=|y(0)+1|$is  the deviation of the tip of the feature from a circle. 
Because there never a cusp in the strict sense of the term the choice of
$\beta_c$,  is a matter of individual judgement. Based on theoretical bubble shapes obtained in Fig. \ref{theorybubble}, $\beta_c$ is chosen to be $0.45$.  

When $R_0  \gg  \kappa/a$, the $x^2$ term in Eq. (\ref{Wulff8}) dominates.  The
boundary around the cusp-like feature is this $x^2$ term instead of the
unperturbed circle.  The new definition of the excluded angle is as displayed in
Fig. \ref{excluded}b and we find $\Psi$ can be expressed as 
\begin{equation}
\Psi\approx 2 \sqrt{2\left|\frac{2(1-\alpha)^2+C_1\alpha}{\alpha^3}
\ln(1-\alpha)\right|}.\label{Wulff16} \end{equation}
In this limit, $\alpha\rightarrow 1-\epsilon$, where
$\epsilon \propto r^{-1}\ll 1$, we approximate $C_1$ as follows
\begin{equation}
C_{1}\approx 2\epsilon(\epsilon-2\delta)+O(\delta\epsilon^2), \label{Wulff13}   
\end{equation}
We then have the large-$R_0$ behaviour of the excluded angle 
\begin{equation}
\Psi\approx 2\sqrt{\frac{8\kappa}{\sigma_0 R_0}\left|\ln\frac{\kappa}{a
R_0}\right|} \label{Wulff17}
\end{equation}

\narrowtext

\subsection{ Polar Coordinates}

As an alternative, one can parameterize the domain boundary in polar
coordinates. In this case, we choose an origin, nominally at the center of
the undistorted circular bubble. The cartesian coordinates $x$ and $y$ are
replaced by $r$, the distance from the origin and $\phi$, the angle with
respect to the $x$ axis. In terms of $r$ and $\phi$, the inverse boojum
texture has the form
\begin{equation}
\Theta \left( r, \phi \right) = \frac{1}{i} \left[ \ln \left( 1 -
\frac{\alpha }{r} e^{i \phi } \right) - \ln \left( 1 - \frac{\alpha
}{r} e^{ -i \phi } \right) \right]. \label{polar1}
\end{equation}
The boundary condition, Eq. (\ref{bound_eqn}) leads to the following result
for the parameter $\alpha$:
\begin{equation}
\frac{ \alpha}{R_0} = \frac{1}{\sqrt{\left( \kappa/\alpha R_0
\right)^2 + 1} + \left(\kappa/\alpha R_0\right)}, \label{polar2}
\end{equation}
where $R_0$ is the radius of the circular bubble. The parameter $\alpha$
has a simple interpretation---it is the distance of the singularity
associated with the virtual boojum from the center of the bubble. Note that
according to Eq. (\ref{polar2}) the singularity always lies within the bubble.

We now proceed with a calculation of the distortion of the bubble from a
circle. Our results are the leading order terms in an expansion in the
dimensionless, and presumably small, combination $a/\sigma_0$. If one writes
\begin{equation}
r(\phi) = R_0 e^{k(\phi)}  ,   \label{polar3}
\end{equation}
one eventually arrives at
\begin{eqnarray}
&&R_0\sigma_0 k^{\prime \prime}(\phi)=\nonumber \\
&&-a \alpha \left[
\frac{\left(e^{i\phi} - \frac{\alpha}{R_0} \right) e^{i \phi}}{\left( 1 -
\frac{\alpha}{R_0} e^{i\phi} \right)^2} + \frac{\left(e^{-i\phi} -
\frac{\alpha}{R_0} \right) e^{-i \phi}}{\left( 1 - \frac{\alpha}{R_0}
e^{-i\phi} \right)^2} \right] \nonumber \\
&&- \frac{2 \kappa \left( \frac{\alpha}{R_0} \right)^2}{\left( 1 -
\frac{\alpha}{\rho} e^{i\phi} \right) \left( 1 - \frac{\alpha}{\rho}
e^{-i\phi} \right)}   .\label{polar4}
\end{eqnarray}

Eq. (\ref{polar4}) represents the leading term in an expansion in the
dimentionless combination $a/\sigma_0$. This ratio is  small, as
the line tension is assumed to be nearly isotropic.

The double integration  of Eq. (\ref{polar4}) is relatively straightforward.
One finds for $k(\phi)$:
\begin{eqnarray}
k(\phi) &=& \frac{2 \kappa}{\sigma_0 R_0}\left[ 2 \int_{-\pi}^{\phi} \arctan
\left( \frac{\frac{\alpha}{R_0} \sin \phi'}{1 - \frac{\alpha}{R_0}
\cos \phi'} \right) d\phi' \right. \nonumber \\
&&- \ln \left( 1 + \left( \frac{\alpha}{R_0} \right)^2 - 2
\frac{\alpha}{R_0} \cos \phi \right)\nonumber \\ 
&&\left. + 2 \ln\left( 1 +
\frac{\alpha}{R_0} \right) \right]  .  \label{polar5}
\end{eqnarray}

\begin{figure}
\makebox[3in]{\epsfbox{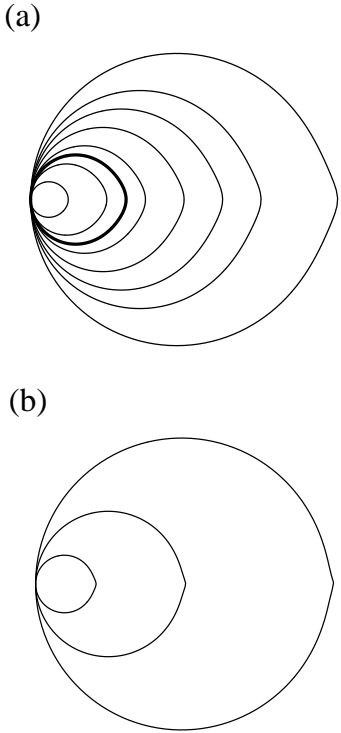}}
\caption{(a) The calculated shapes for the bubble of $R_0$=0.4, 0.8, 1, 1.2, 1.6, 2, 2.4 and 3.2
$\mu m$. The bubble which is drawn in thicker line is the one of the nominal size, $R_c=1 \mu
m$ when $\beta_c=0.45$ for this case.  (b) The calculated shapes for the bubbles of $R_0$=3.2,
8 and 16 $\mu m$. The parameters used are $\kappa/a=0.4\mu m
$ and $a/\sigma_0=0.16$.}
\label{theorybubble}
\end{figure}

In the regimes in which comparisons can be made, the results of the calculation
based on the utilization of polar coordinates  are in complete accord with
results that follow from the Wulff construction. In addition, an analysis
based on polar coordinates---in particular Eq.( \ref{polar5})---provides figures
that can be utilized for comparison with experimental data.  Bubbles whose
shapes are described by Eq. (\ref{polar3}) and Eq. (\ref{polar5}) are displayed in
 Fig. \ref{theorybubble}. These bubbles comprise the results  of theoretical
modeling that are compared with the data described in earlier sections.

\section{Results and conclusions}
We now compare the theoretical excluded angles with the experimental ones. 
Within the range of experiment data, Eq. (\ref{Wulff11}) provides the basis for
comparison between experiments and theory. We find that the best fit is achieved
when the parameters are chosen as $\kappa / a = 0.4 \mu m$ and $a/\sigma_0 = 0.16$.
The comparison between theory and experiment is displayed in Fig.~\ref{exptheory}. 
Because of the fact that there is never a cusp in the form
of a discontinuity in the slope of the bounding curve, we are unable to
\vadjust{\vskip 0.1in
\begin{figure}
\makebox{\epsfbox{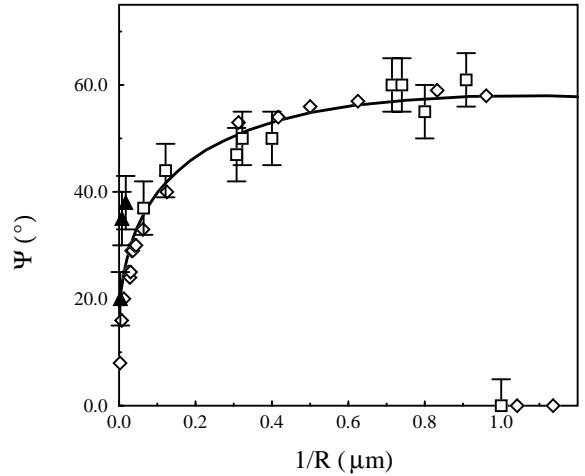}}
\caption{The experimental measured excluded angle of the cusps in bubbles, as in 
Fig. 6, shown as empty squares and filled triangles,  the  measured excluded angle
of the calculated bubbles shown as empty diamond and the analytic expression of
the theoretical excluded angle shown as solid line,  plotted vesus the
reciprocal of the bubble radius. Note the fundamental difference between the
behavior of the  the small-radius behavior of the theoretical curve, which
tends monotonically to a finite limit at vanishing bubble radius and the
measured excluded angle of calculated bubbles, which seems to rise rapidly
from zero as the bubble radius exceeds  a ``threshold'' value. } 
\label{exptheory} 
\end{figure}
\vskip 0.05 in}
definitively model the ``onset'' of a cusp. However, we do find an apparent
rapid dropoff to zero of the (nominal) excluded angle. Furthermore, the theory
predicts that sufficiently small  bubbles  will be {\em nearly} circular. It is
possible to determine a value of what appears to be the radius of onset,
$R_c$, with the exercise of a little judgement. For the parameters of choice, we
find $  R_c =1\mu m$ for $\beta_c=0.45$.

Our theory appears to reproduce fairly well both the qualitative and the
quantitative  behaviour of  the bubble shapes as a function
of the bubble size.  Small bubbles are nearly circular and  the
excluded angle, $\Psi$, is equal to zero. Cusp-like features are observed
for bubbles having  radii $R_0$ greater than  a threshold value  $R_c$. The
angle  $\Psi$ decreases for bubbles of larger sizes. However, in the
experiments,the excluded angle $\Psi$ tends to a non-zero
constant as $R_0 \rightarrow \infty$.  This limiting behavior is not achievable
within the framework of the theory presented here. It is known that thermal
fluctuations will cause the line tension coefficients to depend on the diameter
of the bubble or domain\cite{HSaleur}. However, preliminary calculations
indicate that this renormalization  has no effect on the asymptotic behavior of
the excluded angle for large domains of bubbles. A possible source of the
discrepancy is the difference between the bend and splay moduli, which has been
neglected in the present work.

The ratio $\kappa/a$ for  hexadecanoic acid  has been obtained  by 
Rivi\`{e}re and Meunier \cite {RivMeun}. They find $28\mu m<\kappa/a<40\mu m$, 
about two orders of magnitude larger from our best estimate,
$\kappa/a=0.4\mu m$. Studies of other textures in monolayers such as
stripes\cite{ruiz} indicate that $\kappa$ increases with chain length. This
could account for some of the difference. Another  possible reason for this
discrepancy is the fact that Rivi\`{e}re and Meunier adopted $\kappa_1 \neq
\kappa_3$ as the basis for texture distortions while we have used $a$,  the
anisotropic line tension, as the mechanism for bubble shape distortions.  Our
comparison between theory and experiment gives an estimate of $a/\sigma_0 =
0.16$ for pentadecanoic acid at $LC$-$G$ boundary. The line tension anisotropy,
which  corresponds  to of our $a/\sigma_0$, has  been
estimated for the $LC$-$LE$ interface in D-myristoyl alanine from studies of
dendritic growth\cite{Akamatsu}. The value found is 0.005 - 0.025. Given the
fact that the systems are not identical, and that these are only estimates of
the anisotropy, the difference is not at all disturbing. 

In summary,  we have formulated a theory for bubbles in
Langmuir monolayers that is an extension and reformulation of the theory of
domain shapes discussed by Rudnick and Bruinsma \cite{RudBru}.  Although the
current theory predicts a smooth bubble boundary,  there exist cusp-like features
with well-defined and measurable excluded angles when the radius of the
bubble exceeds a nominal value $R_c$.  Comparing theoretical excluded angles 
with  the results of experiments, we find good qualitative and quantitative
agreement. However,  there are also discrepancies.  Experimental observations
and related work indicate that the difference between $\kappa_1$ and $\kappa_3$
cannot be neglected. 
The calculation of an equilibrium texture when $ \kappa_1 \neq \kappa_3$ is
currently being performed as a prelude to the determination of the shape of a
bubble or domain.

%
%

%
%

\end{multicols}

\begin{references}
\bibitem{PFM} V. T. Moy, D. J. Keller, H. E. Gaub and H. M. McConnell, {\em J.
Phys. Chem.} {\bf 90}, 3198 (1986); D. K. Schwartz and C. M. Knobler, {\em J.
Phys. Chem.} {\bf 97}, 8849 (1993).
\bibitem{BAM} S. H\'{e}non and J. Meunier, {\em Rev. Sci. Instrum.} {\bf 62},
936 (1991); D. H\"{o}nig and D. M\"{o}bius, {\em J. Phys. Chem.} {\bf 95}, 4590
(1991).
\bibitem{smectics} T. Fischer, C. M. Knobler and R. Bruinsma, {\em Phys. Rev.
}{\bf E 50}, 423 (1994).
\bibitem{review} For a description of monolayer phases see C. M. Knobler  and
R.C. Desai, {\em Ann. Rev. Phys. Chem.} {\bf 43}, 207 (1992).
\bibitem{L&S} S. A. Langer and J. P. Sethna, {\em Phys. Rev. A} {\bf 34}, 5035
(1986). For the original discussion of boojums, see N. D. Mermin in {\bf Quantum
Fluids and Solids}, S. B. Trickey, E. Adams and J. Duffy, eds. (Plenum, New
York, 1977).
 \bibitem{Wulff} G. Wulff, {\em Z. Kristalogr.} {\bf 34}, 449 (1901); M. Wortis
in {\em Chemistry and Physics of Solid Surfaces VII}, edited by R. Vanselow and
R. Howe (Springer Verlag Berlin, 1988), p. 367.
\bibitem{RudBru} J. Rudnick and R. Bruinsma, {\em Phys. Rev. Lett.} {\bf 74},
2491 (1995).
\bibitem{Schwartz} D. K. Schwartz, M.-W. Tsao and C. M. Knobler, {\em J. Chem.
Phys.} {\bf 101}, 8258 (1994).
\bibitem{RivMeun} R. Rivi\`{e}re and J. Meunier, {\em Phys. Rev. Lett.} {\bf
74}, 2495 (1995). 
\bibitem{GalFour} P. Galatola and J. B. Fournier, {\em Phys. Rev. Lett.} {\bf
75}, 3297 (1995). 
\bibitem{Fischer} B. Fischer, M.-W. Tsao, J. Ruiz-Garcia, T. M. Fischer and C.
M. Knobler, {\em J. Phys. Chem.} {\bf 98}, 7430 (1994).
\bibitem{polyeth} L. F. Chi, M. Anders, H. Fuchs, R. R. Johnston and
H. Ringsdorf, {\em Science} {\bf 259}, 213 (1993).
\bibitem{dimyr} J. M. Mikrut, P. Dutta, J. B. Ketterson and R. C. MacDonald,
{\em Phys. Rev. B} {\bf 48}, 14479 (1993)  .
\bibitem{octa} J. Y. Fang and C. M. Knobler, {\em J. Phys. Chem.} {\bf 99},
10425 (1995).
\bibitem{flow1} S. Schwiegk, T. Vahlenkamp, Y. -Z. Xu and G. Wegner, {\em
Macromolecules} {\bf 25}, 2513 (1992).
\bibitem{flow2} J. Y. Fang, Z. H. Lu, G. W. Min, Z. M. Ai,Y. Wei and P. Stroeve,
{\em Phys. Rev. A} {\bf 46}, 4963 (1992).
\bibitem{trip} B. G. Moore, C. M. Knobler, S. Akamatsu and F. Rondelez, {\em
J. Phys. Chem.} {\bf 94}, 4588 (1990). 
\bibitem{Tsao} M.-W. Tsao, T. M. Fischer and C. M. Knobler, {\em Langmuir},
{\bf 11}, 3184 (1995).
\bibitem{Poulin} P. Poulin, H. Stark, T. C. Lubensky and D. A. Weitz, preprint.
 \bibitem{forthcoming} K. K. Loh and J. Rudnick, in
preparation. 
\bibitem{Burton} W. K. Burton, N. Cabrera and F. C. Frank, {\em
Philos. Trans. Roy. Soc.} {\bf 243}, 291 (1951).
\bibitem{Santesson} L. Santesson, T. M. H. Wong, M. Taborelli, P.
Descouts, M. Liley, C. Duschl, H. Vogel, {\em J. Phys. Chem.} {\bf 99}, 1038
(1995).
\bibitem{HSaleur} H. Saleur and P. Fendley, private communication.
\bibitem{ruiz} J. Ruiz-Garcia, X. Qiu, M.-W. Tsao, G. Marshall and C. M.
Knobler, {\em J. Phys. Chem.} {\bf 97}, 6955 (1993). 
\bibitem{Akamatsu} S. Akamatsu, O. Bouloussa, K. To and F. Rondelez, {\em Phys.
Rev. A} {\bf 46}, 4504 (1992).

\end{references}
\end{document}